\documentclass[%
 reprint,
superscriptaddress,
 amsmath,amssymb,
 aps,
prl,
]{revtex4-2}
\usepackage{graphicx}
\usepackage{dcolumn}
\usepackage{enumerate}
\usepackage{color}
\DeclareGraphicsExtensions{.eps,.ps,.pdf}
\usepackage{amsmath,amssymb,bbold,bm}
\usepackage[utf8]{inputenc}
\usepackage{gensymb}
\usepackage{upgreek}
\usepackage{hyperref}
\usepackage{tabularx}

\begin{document}

\title{Evidence for a square-square vortex lattice transition in a high-$T_\textrm{c}$ cuprate superconductor} 

\author{D.J. Campbell}
\affiliation{LNCMI-EMFL, CNRS UPR3228, Univ. Grenoble Alpes, Univ. Toulouse, Univ. Toulouse 3, INSA-T,  Grenoble and Toulouse, France}
\author{M. Frachet}
\affiliation{LNCMI-EMFL, CNRS UPR3228, Univ. Grenoble Alpes, Univ. Toulouse, Univ. Toulouse 3, INSA-T,  Grenoble and Toulouse, France}
\author{S. Benhabib}
\affiliation{LNCMI-EMFL, CNRS UPR3228, Univ. Grenoble Alpes, Univ. Toulouse, Univ. Toulouse 3, INSA-T,  Grenoble and Toulouse, France}
\author{I. Gilmutdinov}
\affiliation{LNCMI-EMFL, CNRS UPR3228, Univ. Grenoble Alpes, Univ. Toulouse, Univ. Toulouse 3, INSA-T,  Grenoble and Toulouse, France}
\author{C. Proust}
\affiliation{LNCMI-EMFL, CNRS UPR3228, Univ. Grenoble Alpes, Univ. Toulouse, Univ. Toulouse 3, INSA-T,  Grenoble and Toulouse, France}
\author{T. Kurosawa}
\affiliation{Department of Physics, Hokkaido University, Sapporo 060-0810, Japan}
\author{N. Momono}
\affiliation{Muroran Institute of Technology, Muroran 050-8585, Japan}
\author{M. Oda}
\affiliation{Department of Physics, Hokkaido University, Sapporo 060-0810, Japan}
\author{M. Horio}
\affiliation{Physik-Institut, Universit\"{a}t Z\"{u}rich, Winterthurerstrasse 190, CH-8057 Z\"{u}rich, Switzerland}
\author{K. Kramer}
\affiliation{Physik-Institut, Universit\"{a}t Z\"{u}rich, Winterthurerstrasse 190, CH-8057 Z\"{u}rich, Switzerland}
\author{J. Chang}
\affiliation{Physik-Institut, Universit\"{a}t Z\"{u}rich, Winterthurerstrasse 190, CH-8057 Z\"{u}rich, Switzerland}
\author{M. Ichioka}
\affiliation{Research Institute for Interdisciplinary Science, Okayama University, Okayama 700-8530, Japan}
\author{D. LeBoeuf}
\thanks{david.leboeuf@lncmi.cnrs.fr}
\affiliation{LNCMI-EMFL, CNRS UPR3228, Univ. Grenoble Alpes, Univ. Toulouse, Univ. Toulouse 3, INSA-T,  Grenoble and Toulouse, France}

\date{\today}

\begin{abstract}
Using sound velocity and attenuation measurements in high magnetic fields, we identify a new transition in the vortex lattice state of La$_{2-x}$Sr$_{x}$CuO$_4$ (LSCO). The transition, observed in magnetic fields exceeding 35~T and temperatures far below zero field $T_c$, is detected in the compression modulus of the vortex lattice, at a doping level $x=p=0.17$. Our theoretical analysis based on Eilenberger theory of vortex lattice shows that the transition corresponds to the long-sought 45$\degree{}$ rotation of the square vortex lattice, predicted to occur in $d$-wave superconductors near a van Hove singularity.
\end{abstract}

\maketitle

The band structure of 2D metals hosts special points, known as van Hove singularities (vHS), where the density of states diverges. When the Fermi level is tuned towards a vHS, interesting phenomena occur in the form of electronic instabilities and strongly correlated electron physics. This is exemplified in twisted bilayer graphene, where proximity to a vHS at certain ``magic angles'' \cite{Yuan_NatComm19} can stabilize unconventional superconductivity \cite{Cao_Nat18} and correlated insulating states \cite{Cao_Nat18_2}. In Sr$_2$RuO$_4$ a vHS can be tuned to the Fermi level with chemical substitution or strain, yielding a dramatic increase of superconducting $T_{\rm c}$ \cite{Hicks_Science14}, and non-Fermi liquid behaviour \cite{burganov16,barber18}. Finally, in high-$T_{\rm c}$ cuprate superconductors, the closing of the notorious pseudogap has been associated with doping through a vHS \cite{Benhabib_PRL15}. 

In 2D type-II superconductors with $d$-wave gap symmetry, the proximity to a vHS results in unique behaviour of the vortex lattice (VL). Generally, at low enough fields, a triangular vortex lattice minimizes the dominant electromagnetic interactions \citep{Kleiner64}. With increasing field, the intervortex distance decreases and anisotropies in the electronic structure and/or superconducting gap become relevant. They produce anisotropic screening currents and field distributions around vortices that modify the symmetry and orientation of the VL with respect to the crystallographic axes. In a system where the $d$-wave gap anisotropy is dominant, such as YBa$_2$Cu$_3$O$_y$ (YBCO), a square VL with nearest-neighbours along the nodal direction is expected \citep{Ichioka99}, as observed with small angle neutron scattering (SANS) \citep{Keimer94,BrownYBCOVortexTransition}. On the other hand, 
in LSCO close to optimal doping, proximity to a vHS induces a significant Fermi velocity ($v_{\rm F}$) anisotropy that in turn stabilizes a square VL with nearest neighbour direction oriented along the Cu-O bond at low field. \citep{GilardiLSCOLatticeTransn,ChangSDWVLLSCO}. 

The presence of a $d$-wave gap and four-fold Fermi velocity anisotropy are the ingredients of a long-standing theoretical prediction.
Using the Eilenberger theory of the VL~\citep{Eilenberger68}, Nakai \emph{et al.} predicted that in a sufficiently large magnetic field, the vortex lattice will rotate 45$\degree$, through a first order phase transition, changing the nearest neighbour vortex alignment from along the minima in $v_{\rm F}$ to the minima in the superconducting gap $\Delta{}$.
It is thus predicted that eventually, as the upper critical field is approached, the superconducting gap anisotropy dictates the square lattice coordination \citep{NakaiReentrantVLTransformn}.

The $d$-wave superconductor LSCO near its doping-induced Lifshifz transition is a promising candidate material to display such physics. However, the unconventional electronic properties that typify high-$T_{\rm c}$ cuprates, call into question the applicability of conventional theories for vortex physics.  The  prediction of a field-induced square-square vortex lattice rotation thus stands as an important test of such theories. Experimentally, this is however a challenging task. Being an extreme type-II superconductor, a large portion of the LSCO field-temperature phase diagram is dominated by a vortex liquid. Scanning tunneling microscopy (STM), traditionally used to image the VL~\cite{Zhou_NatPhys13}, is not applicable to LSCO as high quality surfaces have not been established. The weak neutron-matter interaction requires SANS experiments to be carried out using DC-magnets to obtain sufficient counting statistics. This fact has prevented SANS to access field strengths of interest. Ultrasound offers a solution to all those challenges: it is sensitive to the elasticity of the VL and can be measured in pulsed magnetic fields, which cover a larger field range.

Here we present an ultrasound study of the VL of LSCO up to 65~T and down to 1.5~K. We find a new elastic anomaly, marked by a sudden jump in sound velocity, just prior to the melting of the VL, at our lowest measured temperatures. Measuring doping levels from $p=0.10$ to $p=0.215$, we see it only at $p = 0.17$ ($T_c = 38$~K). Moreover, the transition is only observed in the in-plane compression mode, leaving other modes unaffected. Given the absence of any sharp electronic phase transition in this temperature-field range, and the fact that the transition is only seen within the VL state, we conclude that the feature can only come from a change in the VL structure. Using Eilenberger equations along with a realistic tight-binding model of the electronic structure of LSCO $p=0.17$, we show that a square-square first order transition is expected in the field and temperature range where the ultrasound anomalies are observed. We hence conclude that the transition detected in ultrasound in LSCO $p=0.17$ is the long-sought square-square VL transformation.

\begin{figure}
    \centering
    \includegraphics[width=0.42\textwidth]{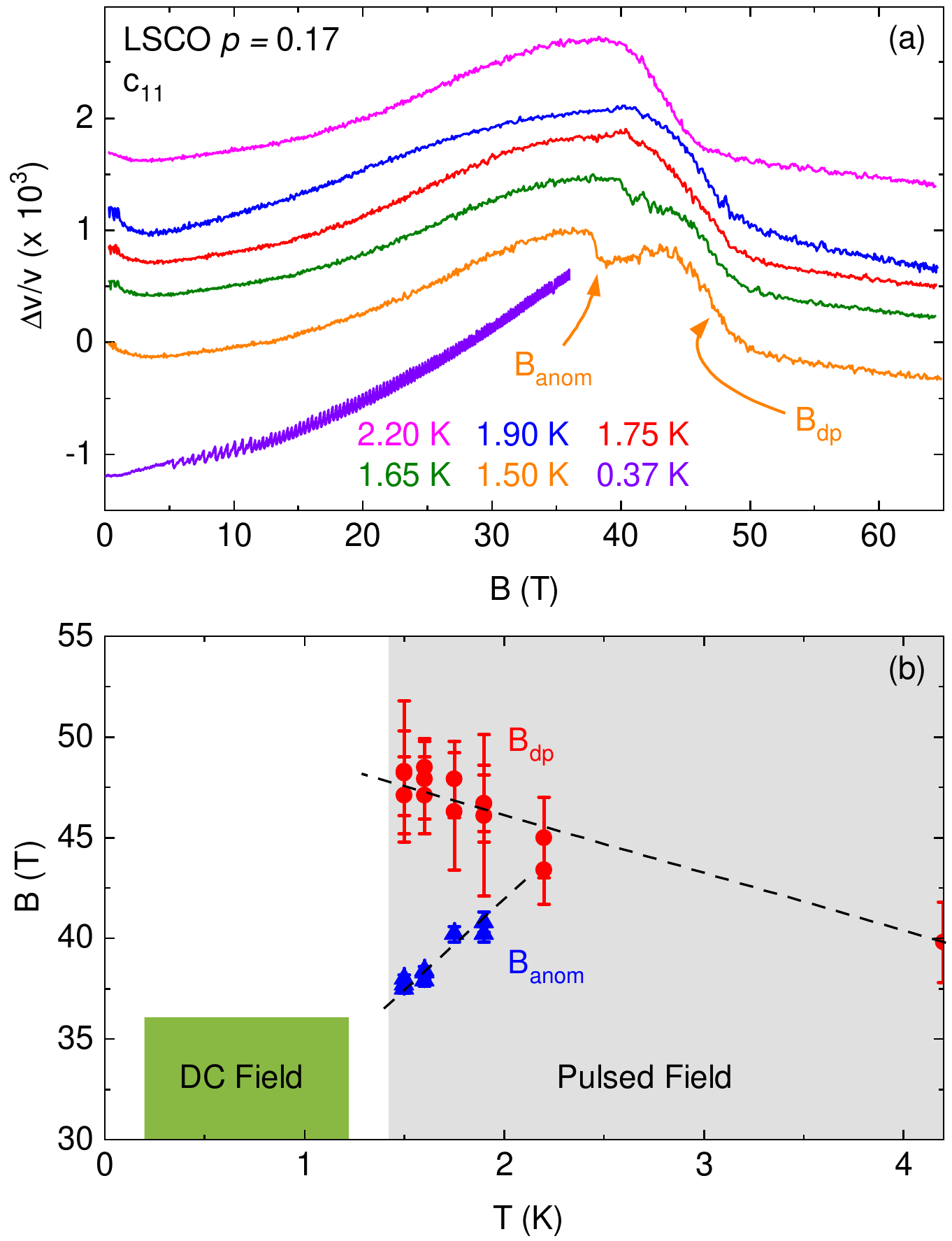}
    \caption{(a) The change in sound velocity of the in-plane longitudinal mode (called $c_{11}$) of a LSCO $p = 0.17$ crystal at different temperatures, as a function of field applied along the \textit{c}-axis. All data are from pulsed field downsweeps, except the lowest temperature curve which is an upsweep in a DC field magnet. This curve shows a serie of spikes above 5 T or so, associated with flux jumps. Curves have been offset vertically (though not by a constant amount). The arrow marks the field of the anomaly, $B_{\textrm{anom}}$, which is visible but more subtle up to 1.90~K in the sound velocity, and the depinning field $B_{\rm dp}$. Measurements were taken at frequencies of 155~MHz in pulsed field and 127~MHz in DC field. (b) Temperature-magnetic field phase diagram, showing the area covered by ultrasound measurements in different magnet systems and the location of the anomaly (blue) and vortex melting (red). Dashed lines are guides to the eye.}
    \label{fig:TDep}
\end{figure}

Data were taken at the two sites of the French National High Magnetic Field Laboratory (LNCMI). Pulsed field measurements up to 65~T and down to 1.5~K were made at the Toulouse site, and DC field measurements up to 36~T and down to 370~mK in Grenoble. Field was always applied along the \textit{c}-axis. The change in sound velocity $\Delta{}v/v$ (equal to $\frac{1}{2}\Delta{}c_{ij}/c_{ij}$ for high symmetry modes, with $c_{ij}$ the corresponding elastic constant) and sound attenuation $\Delta{}\alpha{}$ were tracked using a pulse-echo technique after gluing LiNbO$_3$ transducers to the sides of polished and oriented samples. Hole doping levels are estimated using the high temperature tetragonal-low temperature othorhombic (HTT-LTO) transition temperature determined with sound velocity \citep{FrachetLSCO}. 

The elastic tensor of the triangular VL found in conventional type-II superconductors is made up of three independent elastic moduli (assuming a field along the $c$-axis): a compression modulus $c^v_L=c^v_{11} - c^v_{66}$, a tilt modulus $c^v_{44}$ and a shear modulus $c^v_{66}$ (the Voigt notation is used, see \citep{Campbell72,Brandt95} for reviews). The shear modulus decreases with increasing field and goes to zero at the melting point of the VL \citep{Labusch67, Labusch69}. On the other hand, $c^v_{L} = \frac{B^2}{4\pi} \frac{dH}{dB}$ and $c^v_{44} = \frac{BH}{4\pi}$ (in CGS units). Since our work focuses on properties well above the lower critical field, we assume that the magnetic induction is nearly equal to the applied field. In the field range considered here ($B\sim \frac{1}{2} B_{\rm c2}$), and for extreme type-II superconductors such as LSCO ($\kappa\approx 100$), the ratio of $c^v_{66}$ to $c^v_{44}$ or $c^v_{11}$ is of order $\frac{1}{10}\kappa^2\approx 10^{-5}$ \citep{Campbell72}. Consequently $c^v_{66}<<c^v_{11}=c^v_{44}\approx B^2/4\pi$. In the following we discuss the VL properties of LSCO $p=0.17$ which has square coordination for $B>0.4$~T \citep{GilardiLSCOLatticeTransn,ChangSDWVLLSCO}. In contrast to the triangular VL, the square VL has four independent moduli and the compression modulus is $c^v_L=(c^v_{11}+c^v_{12})/2$. The different elastic constants of the square and hexagonal VL probed for different propagation and polarization directions used here are listed in Table \ref{tab:mode}.

\begin{table*}
    \begin{center}
    \caption{Elastic constants of the crystal lattice ($c^c_{ij}$) measured for different propagation direction $\textbf{k}$ and polarization $\textbf{u}$, in the tetragonal phase. The corresponding elastic constants for the hexagonal and square vortex lattices for applied magnetic field $B\parallel$ [001] are given. For practical purposes, in the text the different modes are referred to as $c_{11}$, L110, $c_{44}$ and T110.}
        \begin{tabular}{|c|c|c|c||c|c|c|}
\hline

mode & \textbf{k} & \textbf{u}&  crystal lattice $c^c_{ij}$ & \multicolumn{3}{c|}{vortex lattice $c^{v}_{ij}$}\\
\hline
&&&  & $\triangle{}_v$ & $\square{}_v$ &  $\square{}_g$ \\

\hline

$c_{11}$ & [100] & [100] & $c^c_{11}$  & $c^v_{11}$ & $c^v_{11}$ & $(c^v_{11}+c^v_{12}+2c^v_{66})/2$ \\
\hline
L110 & [110] & [110] &  $(c^c_{11}+c^c_{12}+2c^c_{66})/2$ & $c^v_{11}$ & $(c^v_{11}+c^v_{12}+2c^v_{66})/2$  & $c^v_{11}$\\
\hline
$c_{44}$ & [100] & [001] & $c^c_{44}$  & none & none & none\\
\hline
T110 & [110] & [1$\bar{1}$0] & $(c^c_{11}-c^c_{12})/2$  & $c^v_{66}$ & $(c^v_{11}-c^v_{12})/2$ & $c^v_{66}$ \\

\hline
    \end{tabular}
    \label{tab:mode}
    \end{center}
\end{table*}  

Pinning of the flux lines is responsible for the coupling between the crystal lattice and the VL. This coupling makes it possible to measure the elastic and pinning properties of the VL with ultrasounds. When the VL is pinned, the sound velocity $v_s$ has a contribution from the VL: $\rho v_s^2=c^c_{ij}+\Delta c^v_{ij}$, with $\rho$ the mass density and $c^c_{ij}$ the crystal lattice elastic modulus. In the early 90's, Pankert \emph{et al.} developed a phenomenological model based on thermally assisted flux flow (TAFF) to describe the influence of the VL on the ultrasound properties in cuprate superconductors \citep{Pankert90a}. The model has been successfully applied to several cuprates, including LSCO \citep{Pankert90, Lemmens91, Hanaguri93}. Within this model, $\Delta c^v_{ij}$ is given by the VL elastic moduli $c^v_{ij}$, renormalized by the VL dynamics associated with the TAFF model.

Data in the $c_{11}$ mode of LSCO $p=0.17$ at 2.2 K (Fig.~\ref{fig:TDep}) illustrate the expected behaviour as a function of magnetic field. In this mode, with field along the $c$-axis, we probe the compression modulus of the VL (see Table \ref{tab:mode}). At low fields, the VL is pinned, such that the sound velocity change follows roughly the $B^2$ increase of the VL compression modulus. At 40 T or so, a step-like decrease is observed in the sound velocity. It corresponds to the depinning of the VL which results in the loss of the vortex contribution to the sound velocity. The ultrasound attenuation $\Delta \alpha(B)$ (Fig.~S1(a), \citep{SI}) shows a broad peak at this depinning field, as also expected within the TAFF model. The broad peak is combined with a monotonic background related in part to the gradual suppression of the superconducting gap \citep{Ikushima64, Kudo03}. 

Below 2~K we see another elastic anomaly emerge, in the form of a smaller but more sudden dip at 1.5~K just below 40~T, and a more subtle change in the signal through 1.9~K (see derivative of $\Delta v/v$ with respect to $B$ in Fig. S1 \citep{SI} where the anomaly is more visible throughout the temperature range). The field at which this feature appears also slightly increases, up to about 43~T. There is a small uptick in the attenuation at the same field \citep{SI}. The new elastic anomaly shows hysteretic behaviour between the up and downsweep of the pulsed field (Fig. S6~\citep{SI}). The phase diagram in Fig.~\ref{fig:TDep}(b) shows the field of this feature ($B_{\textrm{anom}}$, defined as the dip in the derivative shown in Fig. S1 \citep{SI}), as well as the vortex depinning line ($B_{\rm dp}$, defined as the peak in $\Delta{}\alpha{}(B)$). Though the trend line would make it seem as if the anomaly should be observed at low field and lower temperature, no similar behavior was seen down to 0.37~K in DC field measurements up to 36~T on the same sample. This surprising observation has a natural explanation that will be given later.

\begin{figure}
    \centering
    \includegraphics[width=0.45\textwidth]{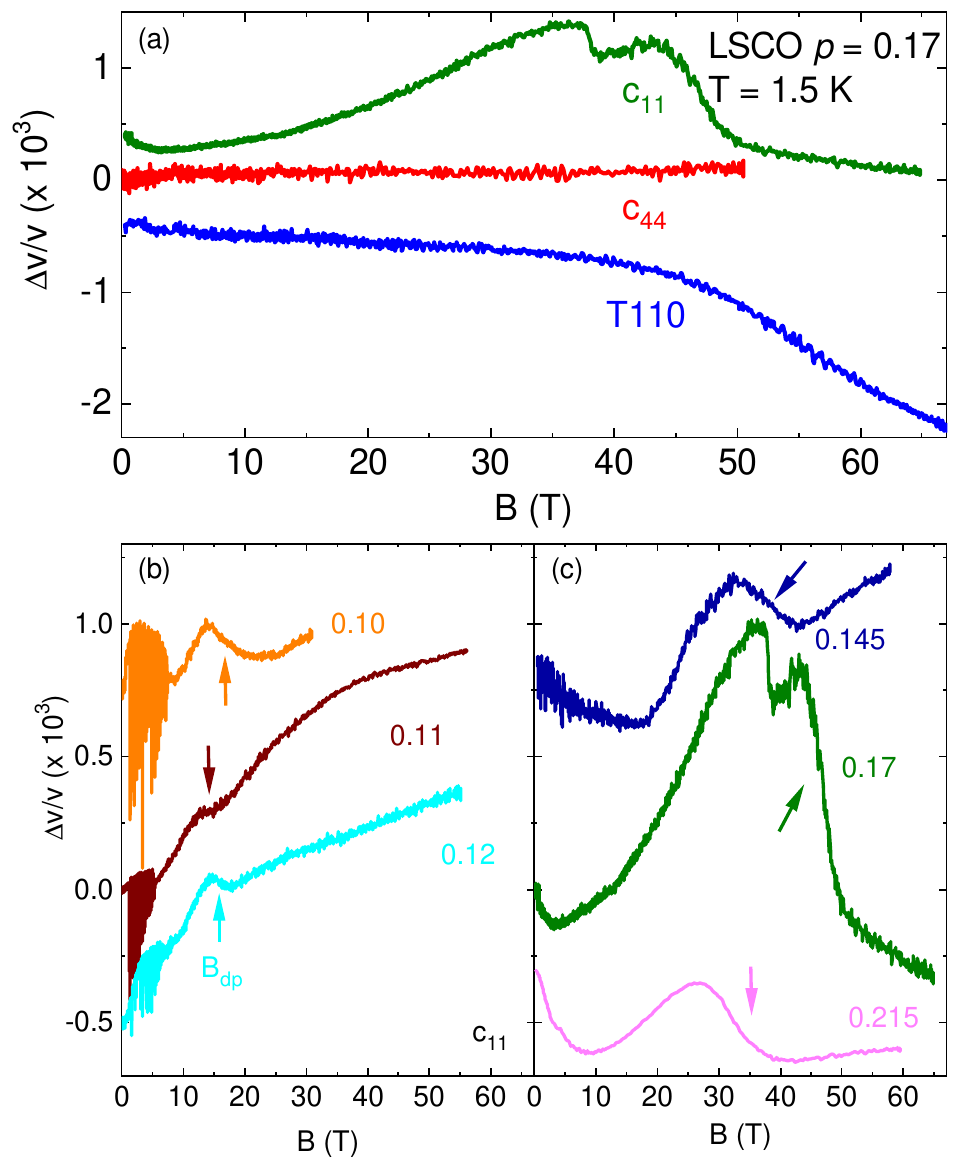}
    \caption{(a) A comparison of the field dependence of three different ultrasound modes of the same LSCO $p = 0.17$ single crystal used in Fig.~\ref{fig:TDep}, taken at 1.5 K, from downsweeps in pulsed field measurements. (b) and (c) Pulsed field data comparing the $c_{11}$ mode for six different LSCO doping levels. Arrows mark the decrease associated with vortex lattice depinning at $B_{\rm dp}$. Note that $p = 0.10$ and 0.11 data are at 4.2~K, all the others are at 1.5~K. The VL contribution appears on top of a background sound velocity, which is associated with the field-dependent electronic properties of LSCO such as superconductivity and magnetism \citep{Frachet21, Nohara95}. This non-vortex background is significant for $p \leq 0.145$ but diminishes smoothly with doping. At low fields, flux jumps appear as a sequence of spikes in the sound velocity of samples with $p\leq 0.145$. Only $p = 0.17$ shows the new elastic anomaly.}
    \label{fig:CombdDep}
\end{figure}

Although we investigated other elastic modes for a similar anomaly, it is only present in $c_{11}$. 
Figure~\ref{fig:CombdDep}(a) shows a comparison of $c_{11}$ to two other configurations at 1.5~K, either for the same sample or one cut from the same original piece. Neither of the other two show an anomaly, and in fact these modes both have a very weak response between low field and 30~T, demonstrating that they are much less sensitive to the vortex lattice. We have probed the L110 mode (see Table \ref{tab:mode}) of LSCO $p = 0.17$, but the signal in this configuration is strongly attenuated below the HTT-LTO structural transition, resulting in a poor signal-to-noise ratio and preventing us from detecting an anomaly (Fig. S2 \cite{SI}). We have also measured the $c_{11}$ mode of other dopings of LSCO [Fig.~\ref{fig:CombdDep}(b)]. All are qualitatively similar, with an increase in the mixed state and dropoff at their corresponding $B_{\rm dp}$. However, the only one with an anomaly before the depinning field is $p = 0.17$, establishing the uniqueness of this doping.

In LSCO $p=0.17$, a spin-glass is present in high magnetic fields; it is characterized by a gradual slowing down of spin fluctuations that produces the smooth softening of the T110 mode (Fig.~\ref{fig:CombdDep}(a)) \citep{FrachetLSCO}. The new sharp feature observed in the $c_{11}$ mode is consequently unlikely to be related to this gradual magnetic freezing. Moreover, the transition is only observed in the in-plane compression mode $c_{11}$, suggesting it is not a structural transition of the crystal lattice. Instead, it is what we expect for a transition in the vortex lattice, as $c_{11}$ is the most sensitive mode to the VL elasticity. It measures the compression modulus of the VL, whereas the $c_{44}$ mode does not couple to the VL, and T110 probes the VL shear modulus $c^v_{66}$ (see Table \ref{tab:mode}), the value of which is well below our resolution. All these observations, along with the fact that the new feature only exists within the pinned vortex solid phase bounded by the red points in Fig.~\ref{fig:TDep}(b), suggest that the new anomaly has its origin in VL physics. 

\begin{figure}
    \centering
    \includegraphics[width=0.45\textwidth]{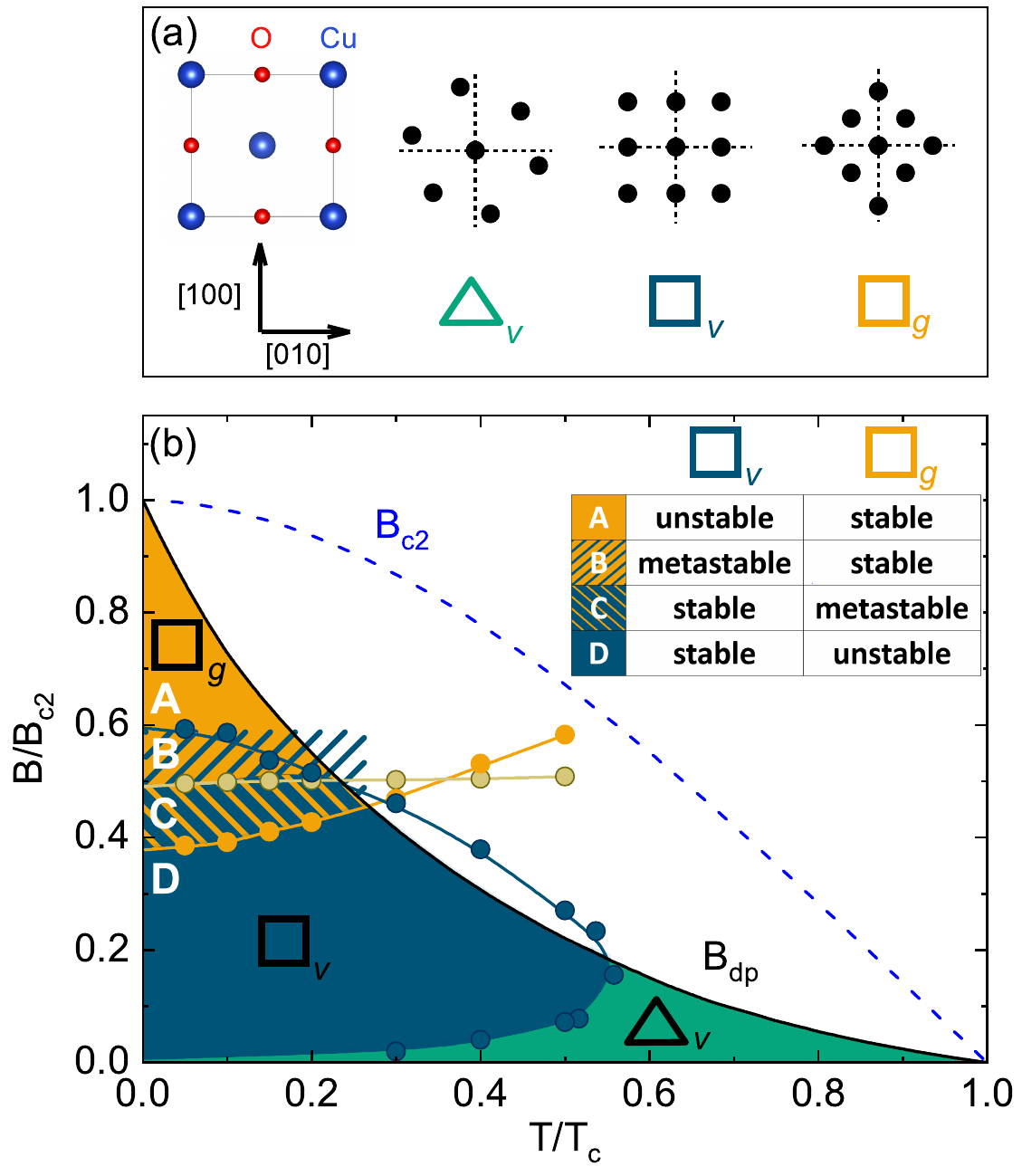}
   \caption{(a) (left) the Cu-O plane \citep{MommaVesta} of LSCO, with the \textit{c}-axis out of the page. (right) Illustrations of the hexagonal, $v_F$-centered square and  superconducting gap-centered square vortex lattices, respectively, relative to the crystal lattice, with vortices portrayed as black circles. Dashed lines mark the Cu-O-Cu bond directions, also the [100] axes of the high temperature tetragonal phase. (b) A phase diagram for $p = 0.17$ showing the calculated VL configurations as a function of $T$ and $B$. The hexagonal phase is stable in the green-colored region at high $T$, low $B$. In regions A and D (orange and blue), one of the two square lattices is stable. The brown line (B-C boundary) marks the transition between the two square VL arrangements in slowly varying magnetic field. However, metastable regions emerge in pulsed fields. Those regions, labeled B and C as indicated in the table in the top right, are shown with slashed lines. In rapid downsweeps of pulsed fields (as shown in Fig. \ref{fig:TDep}), the transition from $\square{}_g$ VL to $\square{}_v$ VL occurs at the C-D boundary, marked by the orange line. In rapidly increasing field, the $\square{}_v$ VL would persist up to the A-B boundary (blue line). The vortex solid is unobservable with ultrasound for $B>B_{\textrm{dp}}$. Solid circles are results of the calculation, lines connecting them are guides to the eye. The depinning transition was not taken into account for the calculations.}
    \label{fig:TheoryPD}
\end{figure}

Since the vortex depinning field is unchanged with the appearance of the feature, an ordered and pinned VL is presumably still present above $B_{\rm anom}$, with only its elasticity or coupling to the crystal lattice being modified. The transition is remarkably sharp, with indication of hysteresis as shown in Fig.~S6~\citep{SI}, pointing to a first order phase transition. The most likely explanation is a structural transformation of the VL. Ultrasound should be sensitive to such transitions, which are expected to involve either a change in elasticity or pinning strength. Indeed, ultrasound has been used to observe a known 45$\degree{}$ rotation of the hexagonal vortex lattice and a rhombic-square transition in YNi$_2$B$_2$C at low fields~\citep{IshikawaYNi2B2C}. In LSCO $p=0.17$, the VL structure goes from hexagonal to square at very low field ($< 0.5 $~T) with nearest-neighbour direction aligned along the Cu-O directions ~\citep{GilardiLSCOLatticeTransn,ChangSDWVLLSCO,DrewLSCOMuSR}. This square VL configuration, presented as $\square{}_v$ in Fig. \ref{fig:TheoryPD}, is at odds with expectations for a $d$-wave superconductor \citep{Ichioka99}, but can be explained by taking into account the anisotropy in Fermi velocity resulting from the proximity to the vHS. The interplay between superconducting gap and Fermi surface anisotropy has previously been investigated with Eilenberger theory \citep{NakaiReentrantVLTransformn}. The calculations showed that at higher fields, further transformations are expected: the $\square{}_v$ can either transition to a hexagonal lattice, or to a gap-centered square lattice $\square{}_g$\textemdash{}the latter being, in essence, a 45$\degree{}$ rotation.

We have performed theoretical calculations following up on Ref.~\citep{NakaiReentrantVLTransformn} based on a 2D tight binding model, with parameters specific to LSCO $p=0.17$ obtained via angle-resolved photoemission spectroscopy (ARPES)~\citep{ChangLSCOAnisotropicScattering,SI}. We evaluate the free energy for a given VL shape, after calculating the vortex state with Eilenberger theory \citep{SI}. Closer to zero field $T_c$ the triangular phase is present, but below about 0.3$T_c$ ($\approx{} 13$~K) only square-square transitions will occur with changing field. From the condition that the free energy increases with a small deformation of the square VL, we can determine the regions (denoted A-D in Fig.~\ref{fig:TheoryPD}(b)) where the $\square{}_v$ and $\square{}_g$ VL are stable. At low temperature and field (regions C and D) the $\square{}_v$ VL is favored, as observed in prior experiment~\citep{GilardiLSCOLatticeTransn}. At higher fields (regions A and B), the contribution of the Fermi surface anisotropy becomes weaker and the influence of the gap structure favors a $\square{}_g$ VL orientation~\citep{NakaiReentrantVLTransformn}. Because the reorientation is a first order transition, VL configurations can have a wide range of metastability within which they can resist temperature or field fluctuations before transitioning to the more energetically favorable configuration~\citep{RLoudenMgB2metastableVL}. Consequently, in regions B and C the unfavored VL configuration ($\square{}_v$ and $\square{}_g$, respectively) can still exist in a metastable state. This means that for downsweeps of pulsed field measurements the $\square{}_g$ VL configuration survives in region C until it undergoes a first order transition to the $\square{}_v$ VL at the C-D boundary (orange circles in Fig. \ref{fig:TheoryPD}). In contrast, in static field measurements, where magnetic field is swept more slowly, the transition is expected at higher field, at the B-C boundary (brown circles).

The theory qualitatively explains the absence of any transition in the DC field measurements. It should only be found at fields above $B_{\rm anom}$ obtained from pulsed field downsweeps. In contrast with Ref.~\citep{NakaiReentrantVLTransformn} which found a re-entrant hexagonal phase at low $T$ between the two square VL configurations, our calculation yields a single, first order transition in LSCO $p=0.17$ at low $T$, as observed experimentally. As noted, a comparison of the up and downsweeps of the field pulse shows evidence for hysteresis (Fig. S6 \citep{SI}), indicating the first order character of the transition. The C-D boundary increases in field with $T$, as does $B_{\rm anom}$. The overall qualitative agreement between theory and experiment is remarkable, and we hence conclude that the ultrasound anomaly at $B_{\rm anom}$ in LSCO $p=0.17$ corresponds to a transition from $\square{}_v$ VL to $\square{}_g$ phase. There is some discrepancy between the transition field predicted by theory (C-D boundary at $B/B_{\rm c2}\approx 0.4$ at low $T$) and the measured $B_{\rm anom}/B_{\rm c2}\approx 0.7$ at 1.5~K, with $B_{\rm c2}=55$~T \citep{FrachetLSCO}. This difference can be explained by at least 2 factors. First, the theoretical curve of $B_{\rm c2}$ may be suppressed at low temperatures if paramagnetic pair breaking and competing electronic orders are considered. Second, including the $c$-axis dispersion in the tight binding model for LSCO may quantitatively shift the C-D boundary~\citep{Horio_PRL18}.

Our data show that the sound velocity in $\square{}_g$ VL is smaller than in $\square{}_v$ VL. This difference can either be due to a smaller rigidity of the $\square{}_g$ VL or enhanced pinning of the $\square{}_v$ VL. Note that we do not observe the hexagonal-to-$\square{}_v$ transition reported using SANS in LSCO for $p\geq0.145$. At $p=0.17$ this transition occurs at $B\approx0.4$~T \citep{ChangSDWVLLSCO}, where the vortex contribution to sound velocity change is small and any anomaly associated with a transition is thus below the measurement resolution $\Delta v/v \sim 10^{-6}$. Our extensive doping dependence (Fig. \ref{fig:CombdDep}) shows that the VL structural transition only exists within a limited doping range around $p=0.17$. This limitation comes partly from the fact that the VL is disordered for $p\leq 0.145$ due to the presence of a spin glass phase that acts as a source of disorder for the VL \citep{ChangSDWVLLSCO}. In LSCO, the VL transition will presumably only exist for doping levels where the spin glass develops at fields larger than the VL transition field, which is the case for $p=0.17$ but not for $p=0.145$ \cite{Chang_PRB08,ChangSDWVLLSCO,FrachetLSCO}. However, while the spin glass phase is not present in LSCO $p=0.215$ \citep{FrachetLSCO}, we do not observe the VL transition at this doping level either (Fig. \ref{fig:CombdDep}). As discussed in detail in Ref. \citep{SI}, the field dependence of the sound velocity at $p=0.215$ in DC magnet shows a broad maximum at low field which is interpreted as a signature of a fishtail effect. The fishtail has been observed in overdoped LSCO using magnetization measurements \citep{Tanabe2ndPeakLSCO}. Its presence signals a disordered vortex state, which cannot support a VL structural transition.


In summary,
we have conducted an ultrasound study of LSCO at several hole concentrations, measuring different elastic modes at low temperature in both pulsed and DC magnetic fields.
For \textit{p}~=~0.17, we observe a novel feature
in the $c_{11}$ mode deep within the mixed phase.
Motivated by theoretical calculations based on Eilenberger equations using realistic tight-binding parameters, we interpret this ultrasound elastic anomaly  in terms of a 45$\degree{}$ square vortex lattice rotation.
Requiring a unique combination of $d$-wave superconductivity and proximity to a van Hove singularity,
this square-square transition is not as widespread as the lower field rhombic-square transition \cite{Zhou_NatPhys13,Keimer94,BrownYBCOVortexTransition,GilardiLSCOLatticeTransn,ChangSDWVLLSCO,IshikawaYNi2B2C}. Even in LSCO, a material that meets both these requirements, the VL transition presumably only exists in a restricted range of the phase diagram, due to the presence of competing electronic phases that act as sources of disorder in the VL. This being said, our results provide experimental evidence of the long standing Eilenberger theory prediction of a magnetic field induced square-to-square vortex lattice transition. 

\section{Acknowledgments}
We thank E. M. Forgan, K. Machida and M.-H. Julien for valuable discussions. Part of this work was performed at the LNCMI-CNRS, a member of the European Magnetic Field Laboratory (EMFL). Work at LNCMI was supported by the Laboratoire d'Excellence LANEF (ANR-10-LABX-51-01), French Agence Nationale de la Recherche (ANR) grant ANR-19-CE30-0019-01 (Neptun) and EUR grant NanoX nANR-17-EURE-0009. M.~H., K.~K., and J.C. acknowledge support by the Swiss National Science Foundation.



\clearpage
\newpage

\onecolumngrid
\appendix

\renewcommand{\figurename}{{\bf Supplementary Figure}}
\renewcommand{\thefigure}{S\arabic{figure}}
\setcounter{figure}{0}

\renewcommand{\tablename}{{\bf Supplementary Table}}
\renewcommand{\thetable}{S\arabic{table}}
\setcounter{table}{0}

\renewcommand\theequation{S\arabic{equation}}
\setcounter{equation}{0}

\setcounter{page}{1}
\section*{Supplementary material for}
\begin{centerline}
{\bf \large Evidence for a square-square vortex lattice transition}
  \end{centerline}
 
 \begin{centerline}
 {\bf \large in a high-$T_\textrm{c}$ cuprate superconductor}
    \end{centerline}

\vskip0.5cm

\begin{centerline}
{D. J. Campbell$^1$ \emph{et al.,}}
\end{centerline}

$^1$LNCMI-EMFL, CNRS UPR3228, Univ. Grenoble Alpes, Univ. Toulouse, Univ. Toulouse 3, INSA-T,  Grenoble and Toulouse, France

\section{Methods}
Measurements up to 36~T were carried out with the resistive magnets of the French National High Magnetic Field Laboratory (LNCMI) site in Grenoble, using a $^3$He refrigerator. These magnets generate a static, ``DC'' field that can be slowly swept or held constant. Higher fields were obtained at the LNCMI's pulsed field site in Toulouse, where measurements were performed down to 1.5~K by pumping on liquid $^4$He. Field was applied along the $c$-axis for all experiments. The change in sound velocity was measured via the pulse-echo ultrasound method using LiNbO$_3$ transducers glued to polished and oriented single crystals. For high symetry modes, the change in sound velocity $v$ directly gives the change in the elastic constant $c_{ij}$ (written with Voigt notation throughout the text) describing those modes, through the basic relation between the two quantities: $c_{ij} = \rho{}v^2$, where $\rho$ is the material density.

\section{LSCO Sample Details}

\begin{table}[!htb]
    \centering
    \caption{Values of hole doping \textit{p}, high temperature tetragonal-low temperature orthorhombic (HTT-LTO) structural transition temperature $T_s$, and superconducting transition temperature $T_c$ for all LSCO samples mentioned in this work. \textit{p} values were determined by the $T_s$ rather than $T_c$, as the former is a monotonic function of $p$ and changes significantly more with doping. However, we used $T_c$ in the case of the $p = 0.215$ sample where no structural transition was observed down to the lowest measured $T$. Within the text, samples have been identified by their values of \textit{p}.}
    \vskip0.5cm
\begin{tabular}{|c| c| c|}
\hline
\textit{p} & $T_s$ (K) & $T_c$ (K) \\\hline
0.10 & $>~275$ & 29~$\pm{}$~2 \\\hline
0.11 & 268 & 26.75~$\pm{}$~3.75 \\\hline
0.122 & 251.8 & 29~$\pm{}$~3 \\\hline
0.149 & 191 & 37.5~$\pm{}$~1 \\\hline
0.17 & 145 & 37.5~$\pm{}$~1 \\\hline
0.215 & $<~7$ & 26~$\pm{}$~2 \\\hline
\end{tabular}
\label{LSCOSamples}
\end{table}

\pagebreak{}

\section{Additional LSCO $p = 0.17$ Data}

\begin{figure}[h]
    \centering
    \includegraphics[width=.6\textwidth]{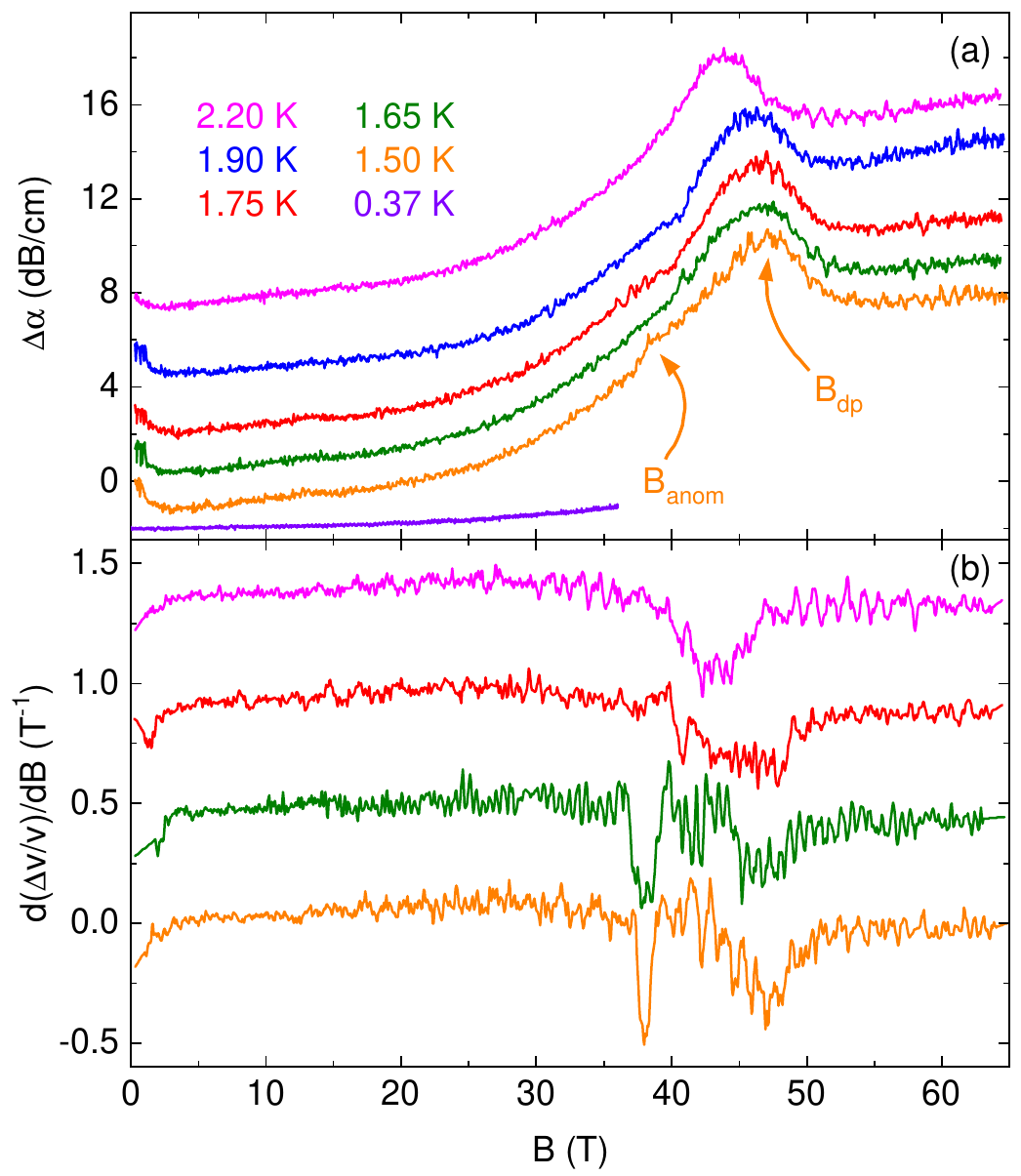}
    \caption{(a) Field dependence of the ultrasonic attenuation in the $c_{11}$ mode of LSCO $p = 0.17$ at low temperature. Arrows mark the field corresponding to the transition in $\Delta{}v/v$ and the vortex lattice melting. (b) The derivative of the change in sound velocity with field for selected temperatures, with the same color coding as the top panel. Offsets have been added to data in both panels. Data were taken at 155~MHz in pulsed field, 127~MHz in DC field.}
    \label{fig:SI-Attn}
\end{figure}

Figure~\ref{fig:SI-Attn}(a) shows the ultrasonic attenuation in LSCO $p = 0.17$, data which were obtained simultaneously with the sound velocity change data seen in Fig.~1(a) of the main text. A subtle increase and change in slope is visible in the 1.5~K attenuation data at the same field as the much more obvious $\Delta{}v/v$ feature. The broad peak in attenuation indicates the vortex lattice depinning.

The lower panel of Fig.~\ref{fig:SI-Attn} shows the derivatives of the sound velocity change with field for selected temperatures shown in Fig~1(a) of the main text. The feature that is more subtle above 1.5~K in those data is clearer here as a dip in the derivative, until 2.2~K where it has either disappeared or been absorbed into the higher field feature corresponding to the vortex lattice depinning.

\begin{figure}[h]
    \centering
    \includegraphics[width=.6\textwidth]{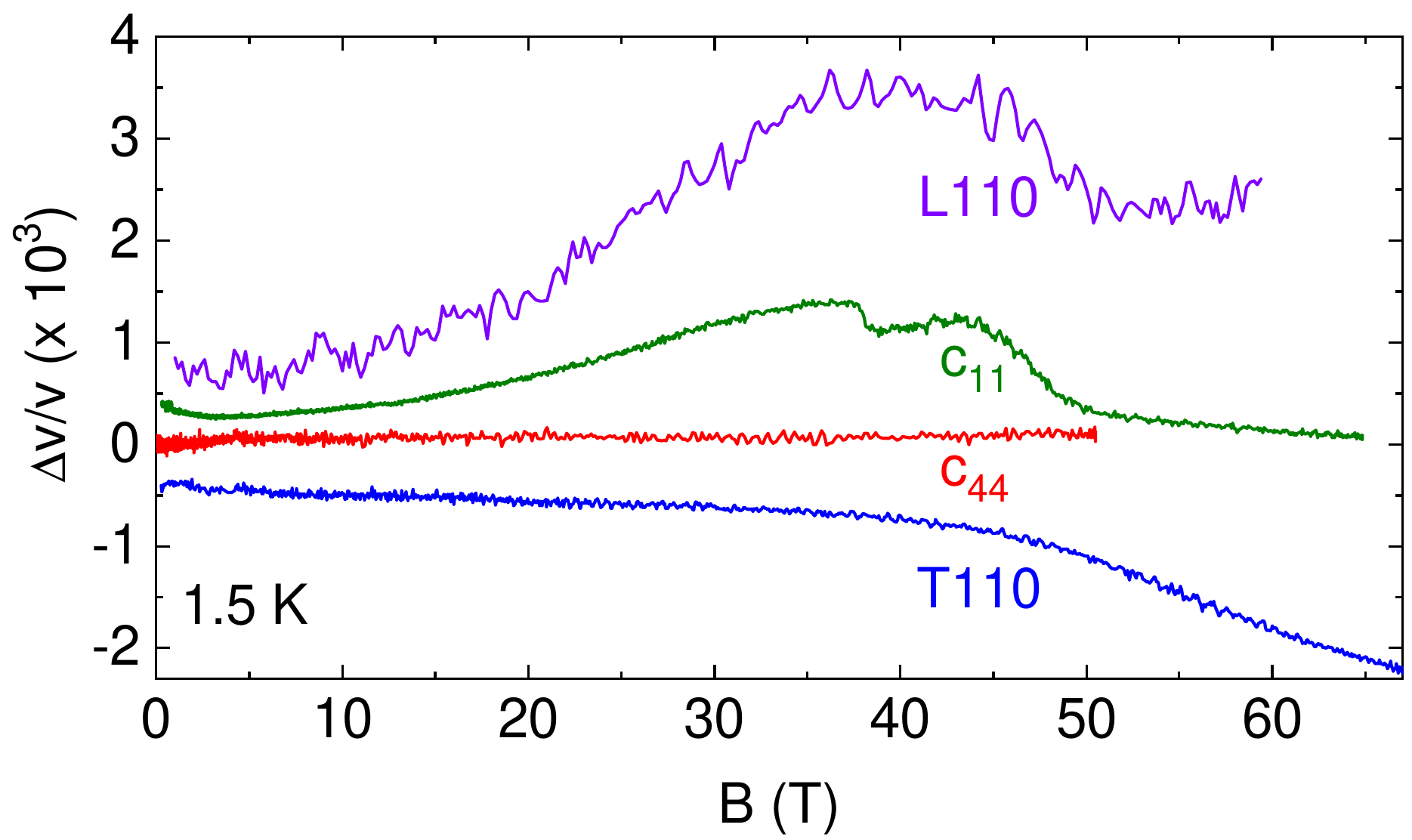}
    \caption{Field dependence of all measured modes of LSCO $p = 0.17$, as shown in Fig.~2(a) of the main text, but now including the L110. This mode is noisier because its signal is significantly reduced in amplitude below the HTT-LTO structural transition.}
    \label{fig:SI-wL110}
\end{figure}

In Fig.~\ref{fig:SI-wL110} we show the results of the longitudinal diagonal mode of LSCO $p = 0.17$ at high field, i.e. the mode with \textbf{u}~$\parallel{}$~\textbf{k}~$\parallel{}$~[110]$_{\textrm{HTT}}$, along with the data shown in Fig.~2(a). This mode corresponds to the $(c_{11}+c_{12}+2c_{66})/2$ elastic constant combination. It has the same qualitative shape as the $c_{11}$ data, but the noise level is such that we cannot pick out a potential transition. The reason for the increased noise is that this mode couples strongly to the HTT-LTO transition that occurs far above base temperature (Table~\ref{LSCOSamples}), and within the LTO phase the signal amplitude is significantly reduced.

\begin{figure}[h]
    \centering
    \includegraphics[width=.65\textwidth]{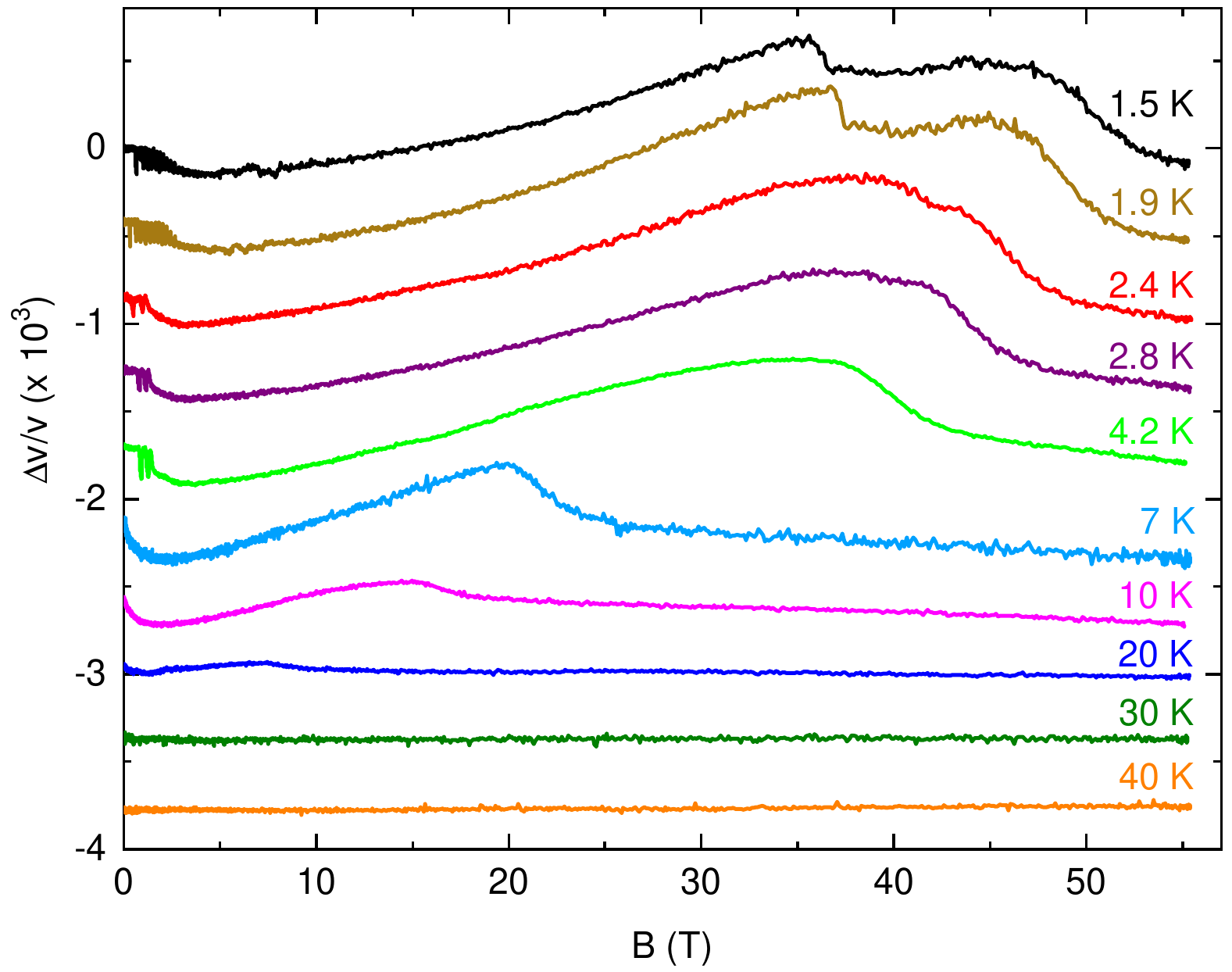}
    \caption{The field dependence of $\Delta{}v/v$ of LSCO $p = 0.17$ sample in the $c_{11}$ configuration, this time extended to higher temperatures than in Fig.~1 of the main text. Curves are separated by a constant offset. Data were taken at 133~MHz. Note that the data set presented in this figure was taken during a different run than the data set presented in Fig.~1(a), but the corresponding $B_{\rm anom}$ have been added to the main text Fig. 1(b).}
    \label{fig:SI-TDep}
\end{figure}

Fig.~\ref{fig:SI-TDep} shows the field dependence of the sound velocity through $T_c = 37.5~\pm{}~1$~K, a much larger temperature range than in the main text. Small humps attributable to the vortex lattice depinning can still be seen at much lower field up to 20~K, but above that there is minimal change with field. 

\pagebreak{}

\section{Calculation Details}

To study potential stable vortex lattice configurations, we estimate the free energy $F$, after calculating the spatial structure of the vortex states by Eilenberger theory. This Eilenberger theory formalism is valid for $\xi_0 k_{\rm F} \gg 1 $, with $k_{\rm F}$ the Fermi wave number and $\xi_0$ the superconducting coherence length. It allows us to quantitatively estimate vortex lattice parameters across the entire $T$ and $B$ range of the vortex state, by evaluating the vortex core structure and the contributions from neighboring vortices. 
  
\begin{figure}[b]
    \centering
    \includegraphics[width=.5\textwidth]{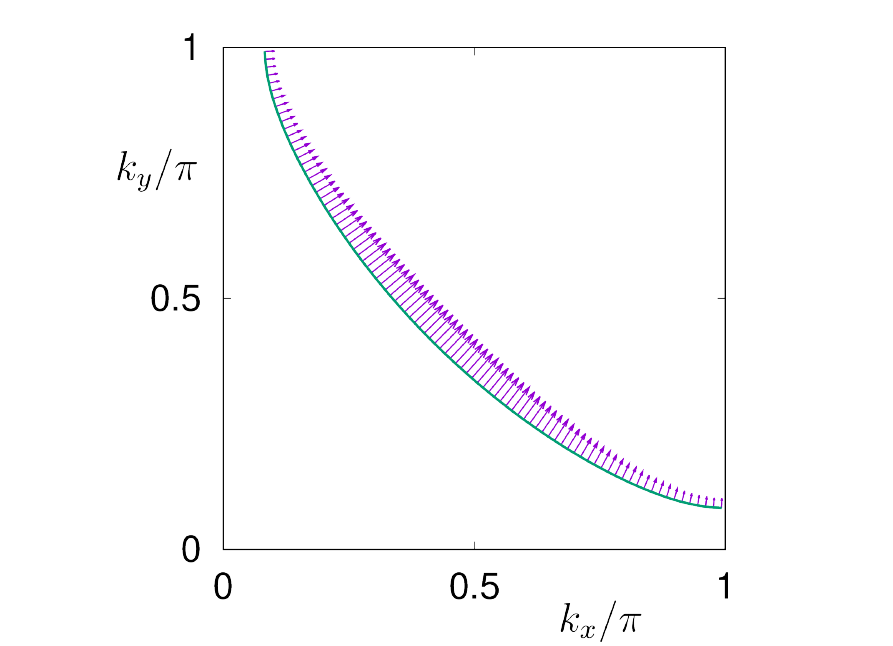}
    \caption{ Fermi surface structure of tight-binding model for LSCO. Lines show the Fermi surface in ${\bf k}$ space. Arrows from the Fermi surface show the magnitude and orientation of the Fermi velocity.}
    \label{fig:TheoryFermiVelocity}
\end{figure}

The Fermi surface of LSCO $p=0.17$  in a two-dimensional tight-binding model is described by the equation  
\begin{eqnarray} 
E(k_x,k_y)=-2t(\cos k_x + \cos k_y) -4t' \cos k_x \cos k_y -2t'' (\cos 2k_x + \cos 2k_y) -\mu =0
\end{eqnarray} 
with $t=0.19$~eV, $t'=-0.027$~eV, $t''=0.013$~eV, and $\mu=-0.144$~eV. The Fermi velocity ${\bf v}=(v_x,v_y)$ is given by 
\begin{eqnarray} 
v_x=\frac{\partial E}{\partial k_x}=(2t + 4t' \cos k_y) \sin k_x +4t'' \sin 2k_x , 
\quad 
v_y=\frac{\partial E}{\partial k_y}=(2t + 4t' \cos k_x) \sin k_y +4t'' \sin 2k_y .
\end{eqnarray}  
The Fermi surface and the Fermi velocity are presented in Fig.~\ref{fig:TheoryFermiVelocity}.  On the Fermi surface, the $d_{x^2-y^2}$-wave pairing function is assumed to be $\varphi({\bf k}) = C( \cos k_y - \cos k_x)$. The normalization constant $C$ is determined so   that $\langle |\varphi({\bf k})|^2 \rangle_{\bf k} =1$, where $\langle \cdots \rangle_{\bf k}$ indicates the Fermi surface average considering ${\bf k}$-dependent density of states $N({\bf k})\propto|{\bf v}|^{-1}$ on the Fermi surface. $N({\bf k})$ is normalized to be $\langle 1 \rangle_{\bf k} = 1$. 	

We assume that a magnetic field is applied to the z-axis, with ${\bf B}=\nabla\times{\bf A}$ and the vector potential ${\bf A}=\frac{1}{2}{\bf H}\times{\bf r} + {\bf a}$,  where ${\bf H}=(0,0,H)$ and $\langle \nabla\times{\bf a}\rangle_{\bf r}=0$. $\langle \cdots \rangle_{\bf r}$ indicates the spatial average within a unit cell of the vortex lattice. In our calculation, the unit cell of the vortex lattice is given by unit vectors ${\bf r}_1=(a_x,0)$ and ${\bf r}_2=(\frac{1}{2}a_x,a_y)$ with $a_x a_y H =\phi_0$, $a_y=\frac{1}{2}a_x \tan \theta$, and the flux quantum $\phi_0$. The triangular (equivalently, hexagonal) lattice with $\theta=60^\circ$ is stable in isotropic $s$-wave superconductors. However, due to the fourfold symmetric structure of the Fermi velocity and the pairing function, the vortex lattice may take on a square, rather than triangular, shape at higher fields.  To study the square lattice $\square_g$ in Fig. 3, we set $\theta=45^\circ$. When we consider the square lattice $\square_v$, we use the Fermi surface rotated by $45^\circ$ in ${\bf k}$ space.   

In a given vortex lattice shape, the spatial structure of the quasiclassical Green's functions $g( \omega_n , {\bf k},{\bf r})$, $f( \omega_n , {\bf k},{\bf r})$, and $f^\dagger( \omega_n , {\bf k},{\bf r})$  at spatial position ${\bf r}$ and the Fermi wave number ${\bf k}$ are calculated in the vortex lattice state by the Eilenberger equations~\cite{Eilenberger68, NakaiReentrantVLTransformn,SuzukiVLOrientation}
\begin{eqnarray} &&
\left\{ \omega_n +\tilde{\bf v} \cdot\left(\nabla+{\rm i}{\bf A} \right)\right\} f=\Delta \varphi g, 
\qquad 
\left\{ \omega_n -\tilde{\bf v} \cdot\left( \nabla-{\rm i}{\bf A} \right)\right\} f^\dagger=\Delta^\ast \varphi^\ast g  , \quad 
\label{eq:Eil}
\end{eqnarray} 
in the clean limit, where $g=(1-ff^\dagger)^{1/2}$, ${\rm Re} g > 0$, $\tilde{\bf v}={\bf v}/v_{{\rm F}0}$ with $v_{\rm F0}=\langle v^2 \rangle_{\bf k}^{1/2}$.  The length, magnetic field, and temperature are in Eilenberger units, i.e. $\xi_0=\hbar v_{\rm F0} / 2 \pi k_{\rm B} T_c$,  $B_0= \phi_0 / 2 \pi  \xi_0^2$, and the transition temperature $T_c$, respectively. The order parameter $\Delta({\bf r})$ and the Matsubara frequency $\omega_n$ are normalized in units of $\pi k_{\rm B} T_{\rm c}$. Equation (\ref{eq:Eil}) is self-consistently solved with the gap equation 
\begin{eqnarray}
\Delta({\bf r}) = g_0N_0 T \sum_{0 < \omega_n \le \omega_{\rm cut}}  \left\langle \varphi^\ast({\bf k}) \left(     f +{f^\dagger}^\ast  \right) \right\rangle_{\bf k} 
\label{eq:scD} 
\end{eqnarray} 
\noindent
and the current equation 
\begin{eqnarray}
\nabla\times \left( \nabla \times {\bf a} \right) =-\frac{2T}{{{\kappa}}^2}  \sum_{0 < \omega_n}  \left\langle \tilde{\bf v}  {\rm Im}~g   \right\rangle_{\bf k}. 
\label{eq:scH} 
\end{eqnarray} 
In Eq. (\ref{eq:scD}), $(g_0N_0)^{-1}=  \ln T +2 T \sum_{0 < \omega_n \le \omega_{\rm cut}}\omega_n^{-1} $   and we use $\omega_{\rm cut}=20 k_{\rm B}T_{\rm c}$. In Eq. (\ref{eq:scH}), we set the GL parameter to $\kappa=100$. The details of the calculation method can be found in Ref. \onlinecite{SeraVortexCalculation}. Using the self-consistent solution, we evaluate the free energy given by~\cite{SuzukiVLOrientation}  
\begin{eqnarray} &&
F={\kappa}^2 \left\langle  |{\bf B}({\bf r})-{\bf H}|^2 \right\rangle_{\bf r} +T  \sum_{|\omega_n|<\omega_{\rm cut}} \left\langle   {\rm Re} \left\langle \frac{g-1}{g+1}(\Delta \phi f^\dagger + \Delta^\ast \phi^\ast f ) \right\rangle_{\bf k}  \right\rangle_{\bf r}  .  \qquad 
\label{eq:f3}
\end{eqnarray} 
in Eilenberger theory. The upper critical field $H_{c2}$ is evaluated as a magnetic field where $\Delta \rightarrow 0$ in the self-consistent solution. At $T \approx{} 0$, $H_{c2} \approx{} 1.3 B_0$.

\begin{figure}[tb]
    \centering
    \includegraphics[width=.3\textwidth]{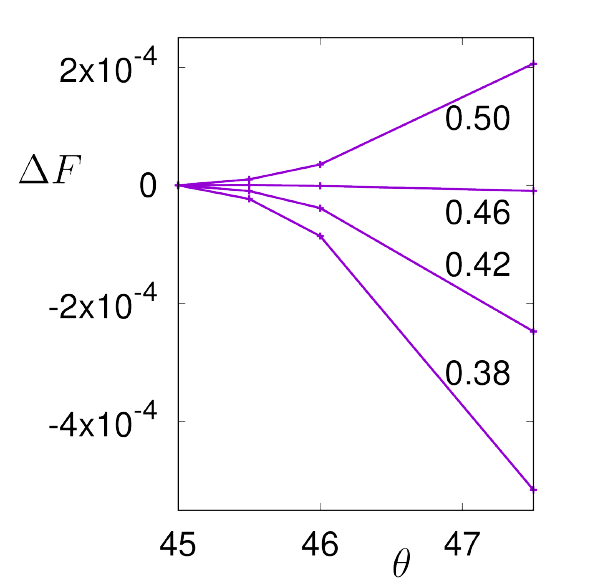}
    \caption{Free energy difference $F_\theta -F_{\theta=45^\circ}$ at $H/H_{c2}= 0.38$, 0.42, 0.46 and 0.50 in the $\square_g$ configuration, with $T=0.2T_c$.  }
    \label{fig:TheoryFreeEnergy}
\end{figure}

For the $\square_g$ configuration,  we evaluate the free energy $F_\theta$ while changing $\theta$ of the vortex lattice deformation, and  show the angle $\theta$-dependence of the free energy difference $F_\theta -F_{\theta=45^\circ}$ at $T=0.2T_c$ in Fig.~\ref{fig:TheoryFreeEnergy}. There, at $H > 0.46$, $F_\theta$ has a minimum at $\theta=45^\circ$, indicating that the square lattice $\square_g$ is stable or metastable. On the other hand, for $H \le 0.46$, $F_\theta$ has a local maximum at $\theta=45^\circ$, and $\square_g$ is unstable  towards deformation to triangular or $\square_v$ lattice. Therefore, by the condition $F_{\square g} \equiv F_{\theta=45^\circ} < F_{\theta=46^\circ} $, we determine the phase of $\square_g$, where the square lattice $\square_g$ is stable or  metastable, in the $T$-$B$ phase diagram of Fig. 3 in the main text. We also determine the phase of $\square_v$ similarly. 

In Fig. 3 of the main text, the $\square_g$ arrangement is possible in regions A, B and C, and the $\square_v$ phase in B, C, and D. While regions B and C can host both $\square_g$ and $\square_v$, $F_{\square g} <  F_{\square v}$ in region B, indicating  that the $\square_g$ vortex lattice is the true free energy minimum, and $\square_v$ is metastable. In region C the situation is reversed: $\square_v$ is stable, and $\square_g$ is metastable since $F_{\square v} <  F_{\square g}$. These results of stability for $\square_v$ and $\square_g$ are summarized in Table~\ref{Theory:Stability}.

\begin{table}[b]
    \centering
    \caption{Stability of square vortex lattice $\square_g$ and $\square_v$ in regions A-D of Fig.~3.  
}
\begin{tabular}{|c| c| c|}
\hline
Region  & $\square_g$ & $\square_v$  \\\hline
A           & stable           & unstable    \\\hline
B           & stable          & metastable  \\\hline
C          & metastable   & stable          \\\hline
D          & unstable       & stable          \\\hline
\end{tabular}
\label{Theory:Stability}
\end{table}

\pagebreak{}

\section{Evidence for hysteresis}
\begin{figure}[h]
    \centering
    \includegraphics[width=.55\textwidth]{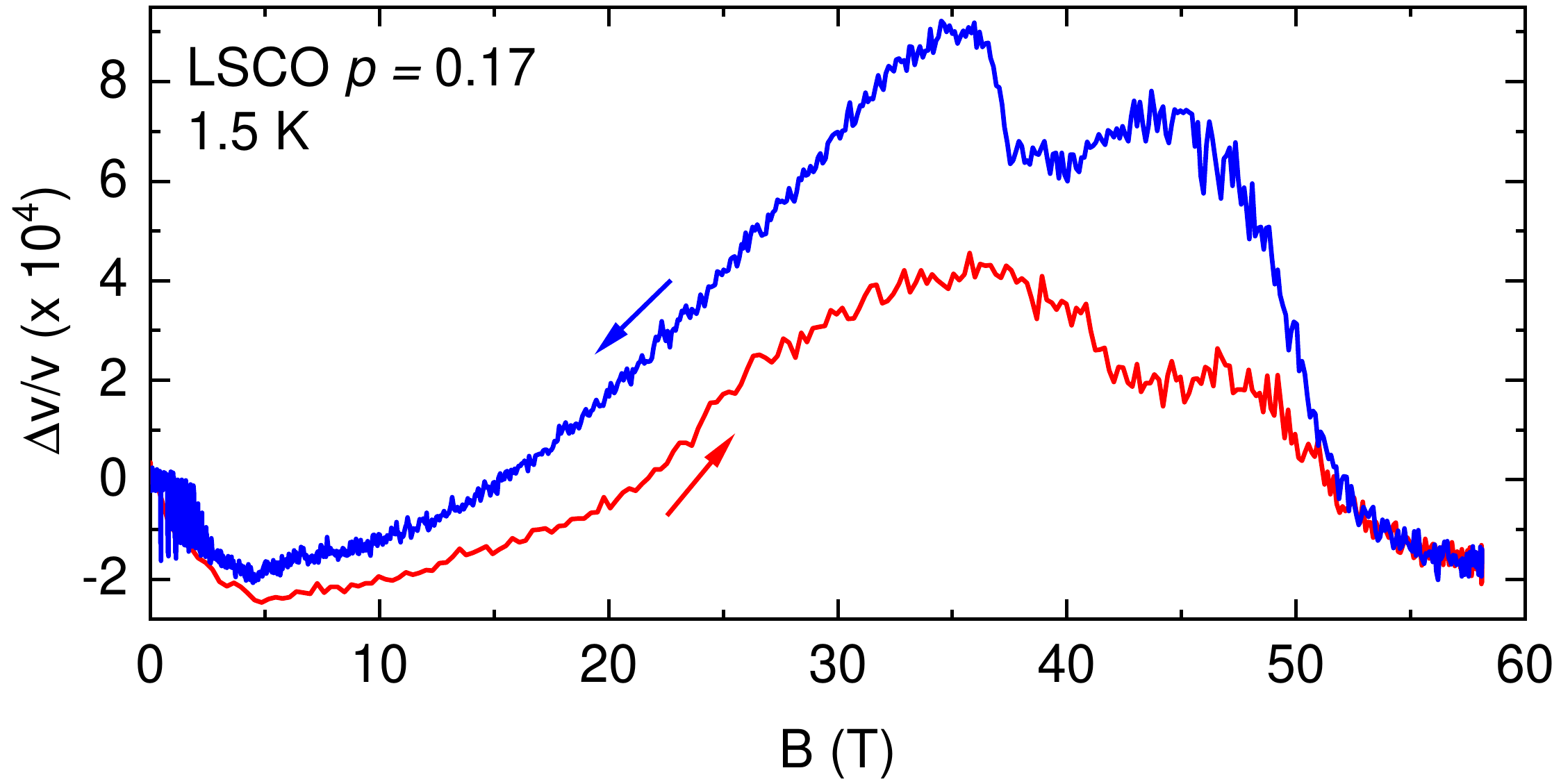}
    \caption{Sound velocity of the $c_{11}$ mode in LSCO $p=0.17$ during a 58 T magnetic field pulse at 1.5 K. The upsweep (downsweep) is shown in red (blue). For increasing fields, the VL transition is found at 41 T while it appears at 37 T or so for decreasing fields. Due to the pulsed-field conditions we cannot rule out eddy current heating of the sample during the pulse. However, the fact that above the irreversibility field the two curves overlap indicates isothermal conditions.}
    \label{fig:SI-hysteresis}
\end{figure}

\section{Second $\Delta{}v/v$ Peak in Overdoped LSCO}

\begin{figure}[h]
    \centering
    \includegraphics[width=\textwidth]{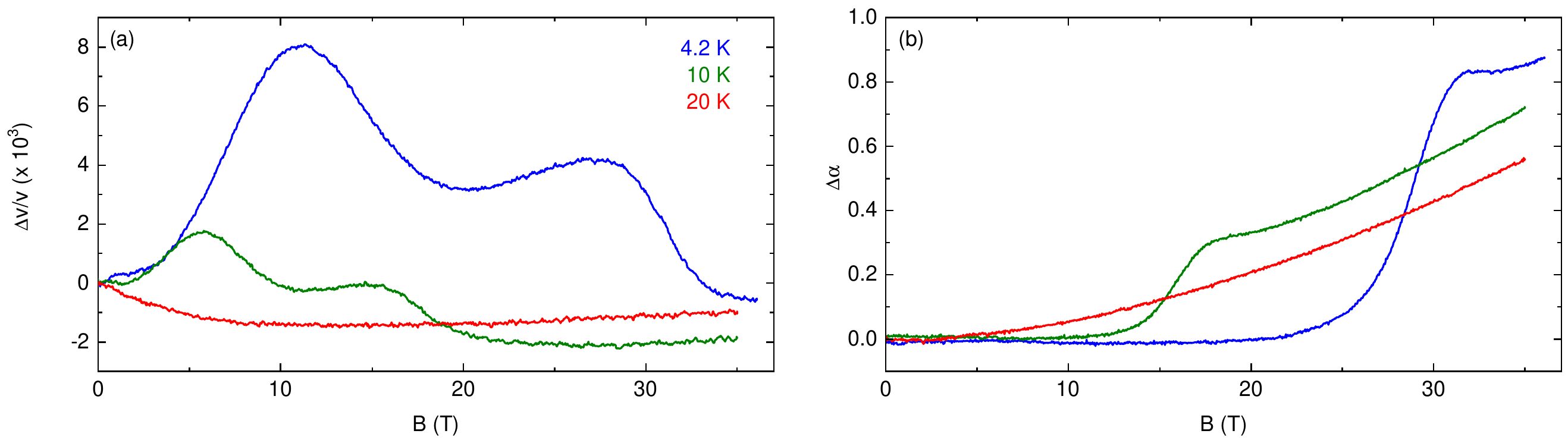}
    \caption{The field dependence of (a) $\Delta{}v/v$ and (b) $\Delta{}\alpha{}$ of the $c_{11}$ mode of a LSCO $p = 0.215$. All data were taken in increasing field.}
    \label{fig:SI-M9}
\end{figure}

In LSCO $p=0.215$, the sharp VL transition is not observed, but measurements of $c_{11}$ in the vortex solid phase show an interesting feature [Fig.~\ref{fig:SI-M9}]. At 4.2 K, a smooth step-like decrease of the sound velocity is found at the depinning field $B_{\rm dp} \approx{} 30$~T, as expected for $c_{11}$. In addition, a broad maximum is found just above 10~T, which we call $B_{\rm peak}$. Similar features are observed at $10$ K. Note that the broad maximum at $B_{\rm peak}$ is not observed in pulsed field data, probably because of the shorter time-scale of those measurements. Nevertheless, a small anomaly could be seen in some pulsed field data sets (for example, at around 5~T and 1.5~K in Fig.~2 of the main text). 
While the sound velocity change at $B_{\rm peak}$ is very similar in appearance to the known $B_\textrm{dp}$ transition at higher field, there is no corresponding feature in the attenuation [Fig.~\ref{fig:SI-M9}(b)]. Therefore, the lower field transition is different from the vortex depinning transition, where there is a sharp rise in the attenuation. The fact that the $\Delta{}v/v$ peak is larger (in DC field), but $\Delta{}\alpha{}$ shows no response, rules out the idea of inhomogeneity in the sample, where different regions could have different $B_{dp}$ values. The presence of a single $T_{\rm c}$ measured with sound velocity in zero field also rules out this scenario. 

\begin{figure}[h]
    \centering
    \includegraphics[width=\textwidth]{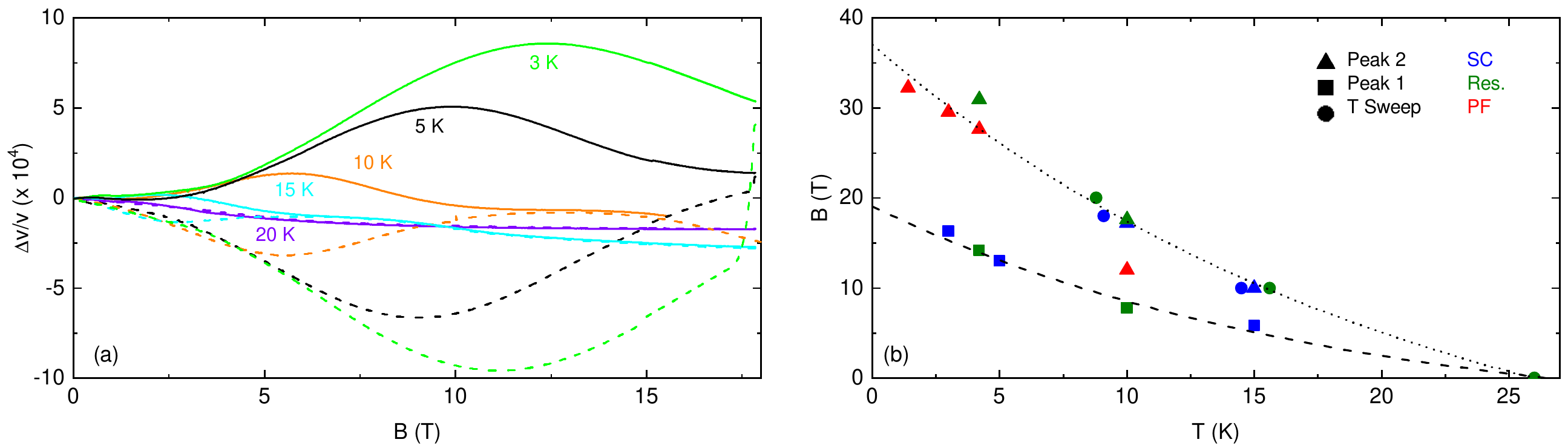}
    \caption{ (a) The field dependence of $\Delta{}v/v$ for LSCO $p = 0.215$ at various temperatures. Data in increasing field are shown as solid lines, in decreasing field as dashed lines. (b) A phase diagram of features in the sound velocity from measurements in superconducting (blue), resistive (green), and pulsed field (red) magnets. Triangles are higher field peaks in sound velocity (judged by the inflection point), squares are lower field peaks, and circles are the inflection points in temperature sweeps at constant field, which correspond to the same irreversibility transition as the triangle symbols. Dashed lines are guides to the eye.}
    \label{fig:SI-M2}
\end{figure}

Data up to 20~T in both increasing and decreasing field also show that $B_{\rm peak}$ is still well below the irreversibility field and that $B_{\rm dp} > B_{\rm peak}$. As seen in Fig.~\ref{fig:SI-M2}(a), $B_{\rm peak}$ is a maximum in $\Delta v/v$ in increasing field, but a minimum at the same field when $B$ is moving toward 0~T. A phase diagram of the inflection points corresponding to the local maximum in $\Delta{}v/v$ shows that the temperature dependence of $B_{\rm peak}$ is qualitatively similar to $B_{\rm dp}$ [Fig.~\ref{fig:SI-M2}(b)], and thus very different from that of the vortex lattice reorientation field $B_{\rm anom}$.

The smooth and broad shape of the feature at $B_{\rm peak}$ suggests it is not a phase transition. A ``second peak'' or ``fishtail'' effect observed in the magnetization of many superconductors, including LSCO, could be the origin of the anomaly at $B_{\rm peak}$. Indeed, the lower field peaks in the sound velocity come at similar field and temperature to those in the magnetization of an LSCO $x = 0.198$ sample~\cite{Tanabe2ndPeakLSCO}. The crystal lattice is has been shown to be sensitive to the fishtail effect via magnetostriction measurements on YBCO~\cite{CuraFishtailMagnetostriction}. In LSCO, the second peak is most evident at overdoping \cite{Tanabe2ndPeakLSCO}, in line with our observation of a second ultrasound peak only in the highest doping we measured. The origin of the magnetization peak is still debated, and in fact there may not be a universal explanation valid for different materials \cite{Miu2ndPeakFeSeTe}. However, we believe that in the case of LSCO $p = 0.215$ it corresponds to a vortex order-disorder transition, possibly to a disordered vortex glass. Enhanced pinning resulting from this process would also cause the observed increase in sound velocity. This has been one of the explanations put forth from measurements on LSCO and other materials~\cite{Tanabe2ndPeakLSCO,Miu2ndPeakFeSeTe}. Increased disorder in the vortex lattice would explain the lack of a VL reorientation transition at this doping-a disordered or glassy vortex lattice, by definition, cannot transition between two ordered structures.

\section*{References}


\bibliography{LSCORefs}


\end{document}